\documentclass[aps,pra,twocolumn,superscriptaddress]{revtex4-1}
\usepackage{graphicx}
\usepackage[colorlinks,urlcolor=blue,citecolor=blue,linkcolor=blue]{hyperref}
\usepackage{color}

\usepackage{stackengine}
\stackMath

\begin{document}

%\preprint{}

\title{Production of a degenerate Fermi-Fermi mixture of dysprosium and potassium atoms}


\author{Cornelis Ravensbergen}
%\thanks{}
\affiliation{Institut f{\"u}r Quantenoptik und Quanteninformation (IQOQI), {\"O}sterreichische Akademie der Wissenschaften, Innsbruck, Austria}
\affiliation{Institut f{\"u}r Experimentalphysik and Forschungszentrum f{\"u}r Quantenphysik, Universit{\"a}t Innsbruck, Austria}

\author{Vincent Corre}
%\thanks{}
\affiliation{Institut f{\"u}r Quantenoptik und Quanteninformation (IQOQI), {\"O}sterreichische Akademie der Wissenschaften, Innsbruck, Austria}
\affiliation{Institut f{\"u}r Experimentalphysik and Forschungszentrum f{\"u}r Quantenphysik, Universit{\"a}t Innsbruck, Austria}

\author{Elisa Soave}
%\thanks{}
\affiliation{Institut f{\"u}r Experimentalphysik and Forschungszentrum f{\"u}r Quantenphysik, Universit{\"a}t Innsbruck, Austria}

\author{Marian Kreyer}
%\thanks{}
\affiliation{Institut f{\"u}r Experimentalphysik and Forschungszentrum f{\"u}r Quantenphysik, Universit{\"a}t Innsbruck, Austria}

\author{Emil Kirilov}
%\thanks{}
\affiliation{Institut f{\"u}r Experimentalphysik and Forschungszentrum f{\"u}r Quantenphysik, Universit{\"a}t Innsbruck, Austria}

\author{Rudolf Grimm}
%\thanks{}
\affiliation{Institut f{\"u}r Quantenoptik und Quanteninformation (IQOQI), {\"O}sterreichische Akademie der Wissenschaften, Innsbruck, Austria}
\affiliation{Institut f{\"u}r Experimentalphysik and Forschungszentrum f{\"u}r Quantenphysik, Universit{\"a}t Innsbruck, Austria}


\date{\today}

\begin{abstract}
We report on the realization of a mixture of fermionic $^{161}$Dy and fermionic $^{40}$K where both species are deep in the quantum-degenerate regime. Both components are spin-polarized in their absolute ground states, and the low temperatures are achieved by means of evaporative cooling of the dipolar dysprosium atoms together with sympathetic cooling of the potassium atoms. We describe the trapping and cooling methods, in particular the final evaporation stage, which leads to Fermi degeneracy of both species. Analyzing cross-species thermalization we obtain an estimate of the magnitude of the inter-species $s$-wave scattering length at low magnetic field. We demonstrate magnetic levitation of the mixture as a tool to ensure spatial overlap of the two components. The properties of the Dy-K mixture make it a very promising candidate to explore the physics of strongly interacting mass-imbalanced Fermi-Fermi mixtures.
\end{abstract}

%\pacs{000}
%\keywords{}

\maketitle

% References should be done using the~\cite, \ref, and \label commands
\section{Introduction}

Strongly interacting systems composed of imbalanced fermions offer an increased complexity as compared to the balanced case by adding effective exchange interaction and dissimilar dispersion of the constituents. The physics emanating from new pairing mechanisms is rather enriched compared to the standard population and mass balanced case, described by the established Bardeen-Cooper-Schrieffer (BCS) attractive interaction \cite{tinkham1996book}. Appearing in various fields, such systems have been predicted to generate exotic pairing such as breached and inhomogeneous pairing, the last exhibiting a spatially varying order parameter \cite{Bennemann2014both}. In nuclear physics the possibility for neutron-proton correlations and Cooper pair condensation has been investigated in several contexts ranging from heavy-ion collisions to astrophysical processes \cite{Brown1994asf}. Inhomogeneous superconductivity has been argued to exist for asymmetric nuclear matter in supernovae and neutron stars \cite{Sedrakian1997sia}. In quantum chromodynamics, a phenomena coined color inhomogeneous superconductivity is expected to take place \cite{Casalbuoni2004isi,Rapp2007csa,Wilczek2007lot,Bailin1984sas,Barrois1977sqm}. Because of its ubiquity such states of matter are also discussed and searched for in condensed matter physics, where experimentally favorable systems include heavy-fermion, organic or high-$T$c superconductors \cite{Casalbuoni2004isi}.
%Exciton condensates in a semiconductor \cite{Yoshioka2011,High2012} as well as exciton-polariton condensate in a semiconductor microcavity \cite{Imamoglu1996}.  

In the context of ultracold two-component Fermi gases some exotic phases like the paradigmatic Fulde-Ferrell-Larkin-Ovchinnikov (FFLO) \cite{Fulde1964sia,Larkin1965iso,Radzihovsky2010ifr}, Sarma \cite{Sarma1963oti} as well as interior \cite{Liu2003igs} and exterior gap breached-pair phases \cite{Gubankova2003bps}, have been theorized. Specifically the FFLO phase seems to be experimentally achievable in mass-imbalanced systems at unitarity, because of the higher superfluid transition temperature \cite{Baarsma2010pam,Gubbels2009lpi,Gubbels2013ifg,Wang2017eeo} relative to the equal- mass case. In general the mass-imbalanced strongly-interacting fermionic mixtures based on cold atoms, as a result of the generated Fermi surfaces and single particle dispersion mismatch, exhibit a rich phase diagram, which includes a Lifshitz point and population dependent asymmetry \cite{Baranov2008spb,Bausmerth2009ccl,Daily2012tot,Diener2010bbc}. Such systems can be also employed to study itinerant ferromagnetism \cite{Massignan2014pdm}, a phenomenon where microscopic understanding is still evolving \cite{amico2018time}. 
%It has been shown that the mass imbalance combined with a narrow Fescbach resonance (FR) extends the lifetime of the repulsive polaron, that eventually limits the domain formation \cite{Massignan2014,Massignan2013}. 
As a further example, a crystalline phase of weakly bound molecules consisting of heavy and light fermions has been predicted \cite{Petrov2007cpo}, countering the intuitive expectation of a gas phase. 

Fermionic mixtures of ultracold atoms are also well suited to tackle impurity physics and the polaron problem \cite{Massignan2014pdm}. One open aspect is an impurity coupled to a non-equilibrium Fermi gas, where multicomponent cold gases are well suited to explore the non-equilibrium Anderson catastrophe and the influence of the environment on the impurity dynamics \cite{Knap2012tdi,you2018atomtronics}. Fermionic mixtures made out of different atomic species additionally allow for the use of a species-selective optical lattices \cite{LeBlanc2007sso}.
% One specific proposal envisions lattice-pinned atoms creating a random potential and thus enabling the observation of Anderson localization \cite{Gavish2005}. 
This handle opens the possibility in multicomponent systems for the realization of Kondo-related physics in transport measurements \cite{Nishida2016tmo}. It should also be emphasized that two-component Fermi gases also offer a variety of interesting few-body effects such as atom-dimer resonant scattering \cite{Levinsen2009ads,Jag2014ooa} and confinement-induced Efimov resonances in systems with mixed dimensionality \cite{Nishida2009cie}.

 Up to now the $^{40}$K-$^{6}$Li mixture has represented the only tunable mixture of fermionic species realized in the laboratory  \cite{Taglieber2008qdt,Spiegelhalder2009cso,Tiecke2010bfr,Ridinger2011lan}. The mass ratio makes this combination in principle attractive for pursuing many of the above goals. In this system, effects of strong interactions near Feshbach resonances (FR) have been observed in hydrodynamic expansion \cite{Trenkwalder2011heo}, in impurity physics \cite{Kohstall2012mac, Cetina2015doi, Cetina2016umb}, and in three-body interactions \cite{Jag2014ooa}. However, at resonant interspecies interaction, this mixture suffers from lifetime limitations owing to the narrow character of the FR \cite{Wille2008eau,Naik2011fri}. The width of the FR, together with Pauli suppression of few-body collisions \cite{Esry2001tlf,Suno2003rot,Petrov2004wbd,Petrov2005spo,Petrov2005dmi,Marcelis2008cpo}, is a prerequisite for achieving long lifetimes in strongly interacting fermionic atomic mixtures and weakly bound dimers made of fermionic atoms. This enhanced stability against inelastic decay has been observed in numerous experiments of both single species~\cite{Dieckmann2002doa,Ohara2002ooa,Cubizolles2003pol,Jochim2003pgo,Baier2018roa} and recently also in dual species experiments \cite{Jag2016lof}. This facilitated the realization and exploration of numerous research avenues, including molecular Bose-Einstein condensation, BEC-BCS crossover physics, the unitary Fermi gas, and superfluid pairing \cite{Pitaevskii2016book,Zwerger2012tbb,Strinati2018tbb}.

% The FR width is also of importance for other research directions, such as the generation of dipolar ground state molecules \cite{Ni2008,Takekoshi2014} or molecular condensates \cite{Jochim2003bec,Greiner2003}.
Here we introduce a new mass-imbalanced Fermi-Fermi mixture, namely the dysprosium-potassium (Dy-K) mixture~\footnote{Mixture experiments with magnetic lanthanide atoms represent a new frontier in the research field. The only other experiment reported so far has been carried out on mixtures of Er and Dy and reported quantum-degenerate Bose-Bose and Bose-Fermi mixtures~\cite{Trautmann2018dqm}}. For creating a Fermi-Fermi mixture, a number of combinations could be selected from the variety of chemical elements that have been brought to Fermi degeneracy~\cite{McNamara2006dbf,Truscott2001oof,Schreck2001qbe,DeMarco1999oof,Naylor2015cdf,DeSalvo2010dfg,Tey2010ddb,Lu2012qdd,Aikawa2014rfd,Fukuhara2007dfg}. The important criteria to adhere to are: (i) mass ratio well below 13.6 to suppress Efimov-related losses \cite{Marcelis2008cpo} (ii) tunable interactions, and (iii) collisional stability \cite{Jag2016lof}. Complying with these criteria narrows down the possible combinations to $^{161}$Dy-$^{40}$K, $^{163}$Dy-$^{40}$K, $^{167}$Er-$^{40}$K and $^{53}$Cr-$^{6}$Li. We have chosen specifically the Dy-K combination, anticipating a favorable scattering spectrum, which is on one hand not chaotic, but it is conveniently dense, extrapolating from \cite{Gonzalezmartinez2015mtf}. We note that, combinations utilizing a closed-shell fermion ($^{171}$Yb,$^{173}$Yb or $^{87}$Sr) in its electronic ground state with an alkali atom offer only extremely narrow resonances~\cite{Zuchowski2010urm,Barbe2018oof}, and therefore are not suited for our purpose.

In this article, we present the preparation and cooling of a Fermi-Fermi mixture of $^{161}$Dy and $^{40}$K atoms, reaching deep quantum degeneracy for both species. In Sec.~\ref{lasercooling}, we first summarize the laser cooling procedures that provide the starting conditions for subsequent evaporation. In Sec.~\ref{evaporation}, we then report our main results. We first demonstrate deep cooling of spin-polarized $^{161}$Dy based on universal dipolar scattering. Then we investigate cooling of a mixture of $^{161}$Dy with $^{40}$K, where the K component is cooled sympathetically by thermal contact with Dy. We also demonstrate the effect of magnetic levitation as an interesting tool for future experiments, and we present a first measurement of the interspecies scattering cross section. In Sec.~\ref{conclusion}, we finally give a brief outlook on future steps to realize a strongly interacting, mass-imbalanced fermionic mixture.

\section{Laser cooling and dipole trap loading\label{lasercooling}}
In this Section, we summarize the laser-cooling steps in our experimental sequence, which prepare the starting point for subsequent evaporative cooling in an optical dipole trap. In Sec.~\ref{MOT} we describe the loading of dysprosium and potassium atoms into two respective magneto-optical traps (MOTs). In Sec.~\ref{sequence} we discuss the procedure used to transfer the atoms into a large-volume optical dipole trap, where they coexist in a mixture.
  
\subsection{Magneto-optical traps\label{MOT}}
Our dysprosium preparation scheme is similar to the one described in Refs.~\cite{Dreon2017oca,Maier2014nlm}. As a source of Dy atoms, we use a high-temperature effusion oven operating at about 1000$^{\circ}$C combined with a Zeeman slower operating on the broad line (natural linewidth of $32\;$MHz) at 421$\;$nm. Dy atoms are collected in a MOT operating on the narrow ($\Gamma_{626}=2\pi\times135\;$kHz) intercombination transition at $626\;$nm. Further details on the apparatus can be found in Appendix~\ref{apparatus}. The MOT uses a magnetic field gradient of about 2$\;$G/cm along the strong vertical axis. We use 35$\;$mm diameter ($1/e^{2}$ intensity drop) beams, with an intensity of $170\;I_{\mathrm{sat},626}$ per beam (where $I_{\mathrm{sat},626}=72\;\mu\mathrm{W\ cm}^{-2}$ is the saturation intensity of the $626\;$nm line). The laser is detuned by a few MHz to the red of the transition~\footnote{We define detunings as $\delta=\omega_{\mathrm{laser}}-\omega_{\mathrm{atom}}$, such that a red (blue) detuned laser corresponds to a negative (positive) detuning} and subsequently spectrally broadened by a high-efficiency electro-optical modulator to increase the loading efficiency~\cite{Maier2014nlm}. Using these parameters we typically load $5\times10^{7}$ atoms of $^{161}$Dy atoms in 3$\;$s. After loading, the MOT is compressed by abruptly switching off the spectral broadening and, within 170$\;$ms, ramping the detuning closer to resonance, the power down to $I\approx 0.5\;I_{\mathrm{sat},626}$, and the gradient of magnetic field down to $1.4\;$G/cm. The atoms are then held in the compressed MOT for 80$\;$ms. During this hold time, the Dy atoms are naturally optically pumped into the stretched state $\vert F=21/2, m_{F}=-21/2\rangle$ \cite{Dreon2017oca}. The temperature after the compressed MOT is approximately $8\;\mu$K.

The source of K atoms is a two-dimensional MOT in a glass cell, connected to the main chamber via a differential pumping tube, similarly to \cite{Ridinger2011lan}. In the chamber the atoms are collected in a three-dimensional MOT operating on the D2 line at a wavelength of $767\;$nm. The K MOT beams combine the cooler and repumper frequencies (the latter being created by a free-space electro-optical modulator). They have an intensity of $I_{\mathrm{cool}}=6\;I_{\mathrm{sat},767}$ and $I_{\mathrm{rep}}=0.3\;I_{\mathrm{sat},767}$, and are detuned by $\delta_{\mathrm{cool}}=-4.4\;\Gamma_{767}$ and $\delta_{\mathrm{rep}}=-2.7\;\Gamma_{767}$ relative to the $F=9/2 \rightarrow F'=11/2 $ and $F=7/2\rightarrow F'=9/2$ transitions, respectively. Here $I_{\mathrm{sat},767}=1.75\;\mathrm{mW\ cm}^{-2}$ is the saturation intensity of the D2 line of K, and $\Gamma_{767}$ its natural linewidth. The gradient of magnetic field is set to $9.6\;$G/cm along the strong axis. We typically load the MOT for $3\;$s, which gives $2\times 10^{6}$ atoms of $^{40}$K. Once loaded, the K MOT is compressed by simultaneously ramping up the gradient of magnetic field to $25\;$G/cm, ramping down the detunings to $\delta_{\mathrm{cool}}=-1.1\;\Gamma_{767}$ and $\delta_{\mathrm{rep}}=0$ and the powers down to $I_{\mathrm{cool}}=0.5\;I_{\mathrm{sat},767}$ and $I_{\mathrm{rep}}\approx 0.01\;I_{\mathrm{sat},767}$,  all in $4\;$ms. At the end of the compressed MOT phase the temperature of the K cloud is $110(30)\;\mu$K. We then perform gray molasses cooling on the D1 line, which allows us to lower this temperature to approximately $30\;\mu$K. Details on the gray molasses stage are given in Appendix~\ref{D1cooling}.

\subsection{Sequential dipole trap loading\label{sequence}}

The transfer of atoms from a MOT or a molasses into an optical dipole trap is a common procedure in cold-atom experiments. The optimum transfer strategy depends on the particular properties of the atomic species and on the cooling scheme applied prior to the dipole trap loading. In many dual-species experiment, however, the optimum loading conditions for both components are incompatible with each other, such that a sequential loading scheme is necessary. In the case of Dy and K atoms, the main constraint is imposed by the vastly different gradient of the magnetic field required in the respective MOTs. Indeed, the larger gradient used to load K atoms would induce a strong compression of the Dy cloud, which in turn would lead to large losses. We therefore sequentially load the K atoms and then the Dy atoms into a reservoir optical dipole trap (RDT).

\begin{figure}
 \includegraphics[width=\columnwidth]{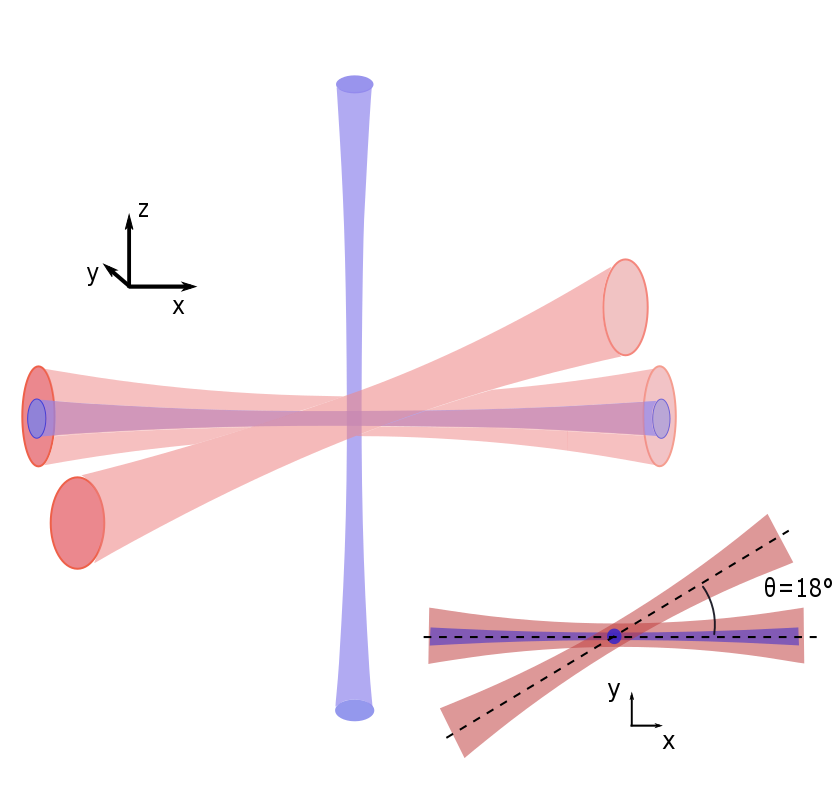}
 \caption{Optical dipole traps scheme. The two beams of the large volume RDT propagate in the horizontal $x$-$y$-plane and intersect under an angle of $18^{\circ}$, while the two orthogonal beams of the CDT propagate in the $x$-$z$-plane. The lower right scheme shows a view from the top.\label{ODTsetup}}
 \end{figure}

The RDT is created by two intersecting beams derived from a single, longitudinally multimode, fiber laser (IPG YLR-100-LP, wavelength $1070\;$nm) and crossing in the horizontal plane under an angle of 18$^{\circ}$; see Fig. \ref{ODTsetup}. The RDT is already switched on during the K-MOT stage, with a power per beam of $11.5\;$W, corresponding to a depth $U_{\mathrm{RDT}}^{\mathrm{K}}/k_{\mathrm{\mathrm{B}}}=260\;\mu$K. After compression of the K MOT (Sec.~\ref{MOT}), the atoms are transferred into the RDT, while being cooled by the gray molasses (App.~\ref{D1cooling}). We typically load $5\times10^{5}$ $^{40}$K atoms, corresponding to a transfer efficiency of 20$\%$, at a temperature of 33(1)$\;\mu$K ~\footnote{In the absence of dipole trap, the D1 cooling allows us to reach temperatures as low as 8(1)$\;\mu$K}. 

After the gray molasses, a $0.8\;$ms laser pulse is applied to optically pump the atoms in the $\vert F=9/2,m_{F}=-9/2\rangle$ state. The pulse has $\sigma^{-}$ polarization with a small admixture of $\pi$ polarization, and has frequency components resonant with the D1 transitions $F=7/2\rightarrow F'=9/2$ and $F=9/2\rightarrow F'=7/2$, such that the $\vert F=9/2,m_{F}=-9/2\rangle$ state is a dark state. During the pulse, the gradient of magnetic field is turned off and a homogeneous field of approximately $1\;$G is applied along the direction of the beam. The optical pumping only has a minor effect on the temperature of the cloud: We measure an increase of the temperature on the order of $1\;\mu$K. To check the polarization of the K sample we perform resonant absorption imaging at high magnetic field ($155\;$G), where the splittings between the different $\vert F,m_{F}\rangle\rightarrow\vert F',m_{F'}=m_{F}+1\rangle$ transitions are large compared with the natural linewidth. After the pulse we still observe approximately $10\%$ of atoms left in $\vert 9/2, -7/2\rangle$. 

We then load the Dy MOT for $3\;$s, while holding the K atoms in the RDT. As we start loading Dy, the power of the RDT is increased, resulting in a depth $U_{\mathrm{RDT}}^{\mathrm{Dy}}/k_{\mathrm{B}}\approx 110\;\mu$K for Dy and $U_{\mathrm{RDT}}^{\mathrm{K}}/k_{\mathrm{B}}\approx 360\;\mu$K for K \cite{Ravensbergen2018ado}. This compression increases the temperature of K to $37(2)\;\mu$K. Since the Dy MOT forms below the zero of the quadrupole magnetic field, it is spatially separated from the K cloud and does not affect the lifetime of the K atoms in the dipole trap, which is measured to be $4\;$s. Finally, the compression of the Dy MOT (Sec. \ref{MOT}) moves it upward, such that it overlaps with the RDT, where Dy atoms are transferred. At this stage the central region of the RDT typically contains $10^{7}$ Dy atoms at a temperature of $17(1)\;\mu$K, corresponding to a loading efficiency of 10$\%$, and approximately $2\times 10^{5}$ K atoms at a temperature of $28(2)\;\mu$K. The fact that the temperature of the K sample decreases during the $3\;$s hold in the RDT is also observed in the absence of Dy and is presumably an effect of plain evaporation allowed by the remaining spin polarization imperfection. We do not observe interspecies thermalization on the considered time scale at this point. 

At the end of the Dy MOT stage, all resonant light is switched off and the gradient of magnetic field is set to zero. A homogeneous magnetic field of approximately $430\;$mG is applied along the vertical direction to define a quantization axis and to maintain the polarization of the two clouds. At this point, we do not observe any population in higher spin states of K and Dy, which we attribute to rapid intraspecies and interspecies dipolar relaxation. The conditions after the loading of both species in the RDT are summarized in Tab.~\ref{evapRDT}.

% \begin{figure}
% \includegraphics[width=\columnwidth]{timeline.png}
% \caption{simplified description of the experimental sequence. \label{timeline}}
% \end{figure}

\section{Evaporative cooling\label{evaporation}}

In this Section, we present our main results on evaporative cooling down to deep degeneracy of both fermionic species. We evaporatively cool fully spin-polarized Dy~\cite{Lu2012qdd, Burdick2016lls} (see also related work on fermionic Er~\cite{Aikawa2014rfd, Aikawa2014ard}), relying solely on universal dipolar scattering \cite{Bohn2009qud}. The K component, for which the optical trap is much deeper, is cooled sympathetically. This cooling scheme is closely related to earlier experiments on the attainment of quantum degeneracy in Yb-Li mixtures \cite{Hara2011qdm} and on Yb-Rb mixtures \cite{Hansen2011qdm, Hansen2013poq}, where the heavy lanthanide species experiences a shallower trapping potential than the light alkali species. As in our work, evaporative cooling acts on the heavier species, which then cools the lighter one in a sympathetic way. The main difference to our work is the origin of elastic collisions among the heavier atoms, which in the Yb cooling scenarios result from $s$-wave collisions between identical bosons or non-identical fermions, and not from universal dipolar scattering between identical fermions. 

We employ a sequence of two optical dipole traps. A first evaporative cooling step is performed in the RDT. Atoms are then transferred into the main dipole trap, where evaporative cooling continues. In Secs.~\ref{Dyalone} and \ref{DyandK}, we characterize the evaporation process in two situations: We first describe the evaporation of $^{161}$Dy atoms alone, and then the evaporation of the mixture. In Sec.~\ref{levitation} we discuss the effect of gravity and its compensation by a levitation field. Finally, in Sec.~\ref{crossedthermalization} we show the results of an interspecies thermalization measurement, from which we give an estimate of the inter-species $s$-wave scattering cross section, a quantity of primary importance in the efficiency of the simultaneous cooling of the two species.

	\subsection{Transfer from reservoir into main dipole trap\label{strategy}}
	
\begin{table}%[H] add [H] placement to break table across pages
 \caption{\label{evapRDT}Cloud parameters after transfer into the RDT. The given parameters describe the cloud in the RDT immediately after the Dy MOT is turned off. $U_{\mathrm{RDT}}$ refers to the depth of the optical potential, $n_{\mathrm{peak}}$ to the peak density in the center of the trap, PSD to the phase-space density. The errors in $n_{\mathrm{peak}}$ and in the PSD results from the propagation of the errors in $N$ and $T$, whereas the systematic errors in $U_{\mathrm{RDT}}$ and in $\bar{\omega}$ are not propagated.}
 \begin{ruledtabular}
 \begin{tabular}{|c|c|c|}
 \centering   & $^{161}$Dy&$^{40}$K \\
 \hline
 $\hspace{0.3cm}\bar{\omega}/2\pi$& $154(6)\;$Hz & $558(22)\;$Hz  \\
 $U_{\mathrm{RDT}}/k_{\mathrm{B}}\hspace{0.3cm}$ & $110(8)\;\mu$K & $360(20)\;\mu$K   \\
   $N$ & $1.2(1)\times 10^{7}$ & $2.0(2)\times 10^{5}$   \\
   $T$ & $17(1)\;\mu$K & $28(2)\;\mu$K\\
   $n_{\mathrm{peak}}$& $2.2(3)\times 10^{13}\;\mathrm{cm}^{-3}$ & $1.2(2)\times 10^{12}\;\mathrm{cm}^{-3}$  \\
PSD& $8(2)\times 10^{-4}$ & $1.7(4)\times 10^{-4}$ 
 \end{tabular}
 \end{ruledtabular}
 \end{table}
	
	 With the phase-space density reaching values in the $10^{-4}$ range for each species, the RDT provides favorable starting conditions for evaporative cooling. However, the large waist of the beams forming the RDT does not allow us to maintain a large enough collision rate in this trap when its power is lowered. Besides this, as a technical issue, the RDT is created by a longitudinally multimode laser, where evaporation of Dy is known to be hindered by heating processes \cite{Maier2015phd,Landini2012dec}. We therefore perform a two-step evaporation: After a first evaporation ramp taking place in the RDT the atoms are transferred into a tighter crossed dipole trap (CDT), where the evaporative cooling continues. 
		
	The CDT is created by a longitudinally single-mode Nd:YAG laser (Mephisto MOPA 18 NE) at $1064\;$nm. The configuration of dipole traps is depicted in Fig.~\ref{ODTsetup}. One beam of the CDT overlaps in the horizontal plane with one beam of the RDT and has a waist of approximately 30$\;\mu$m, while the second beam propagates along the vertical axis and has a waist around 60$\;\mu$m. 
	
	During the first $20\;$ms after completion of the RDT loading, the power of the RDT is kept fixed while the powers in the horizontal and vertical beams of the CDT are ramped up from 0 to $1.6\;$W and $0.7\;$W, respectively. This results in a depth of approximately 54$\;\mu$K for Dy, and of $174\;\mu$K for K. From that point, as illustrated in Fig.~\ref{evapsequence}, the power of the RDT is reduced in an exponential ramp to $1.6\;$W per beam in $100\;$ms, while the power in the horizontal beam of the CDT is ramped up to $3.5\;$W. The power in the vertical beam is kept constant during this time. At the end of these ramps, the RDT is turned off. At its full power, the CDT has a depth $U_{\mathrm{CDT}}^{\mathrm{Dy}}/k_{\mathrm{B}}=130\;\mu$K and the trap frequencies are $[\omega_{x},\omega_{y}, \omega_{z}]=2\pi\times[88,1040,1037]$\ Hz for Dy. Typically, $2\times 10^{6}$ Dy atoms and $6.5\times 10^{4}$ K atoms are loaded into the CDT, with a temperature of $24(1)\;\mu$K for Dy and $28(1)\;\mu$K for K. The phase-space density after the transfer is on the order of $10^{-3}$ for both Dy and K. The cloud parameters immediately after transfer into the CDT, i.e. the starting conditions for the subsequent evaporative cooling, are summarized in Tab.~\ref{evapCDT}.
	
\begin{table}%[H] add [H] placement to break table across pages
 \caption{\label{evapCDT}Starting conditions for the evaporation in the CDT. $N_{\mathrm{center}}$ gives the number of atoms in the central region of the trap, while $N_{\mathrm{arms}}$ corresponds to the atom number trapped outside the region where the two beams cross, in the arms of the potential. $U_{\mathrm{CDT}}$ refers to the depth of the optical potential created by he CDT, $n_{\mathrm{peak}}$ to the peak density and PSD stands for phase-space density. As in Table~\ref{evapCDT}, the errors in $n_{\mathrm{peak}}$ and in the PSD is obtained from the errors in $N$ in $T$ only.}
 \begin{ruledtabular}
 \begin{tabular}{|c|c|c|}
 \centering
  & $^{161}$Dy & $^{40}$K \\
 \hline
  $\bar{\omega}/2\pi$& $456(18)\;$Hz  & $1460(60)\;$Hz \\
 $U_{\mathrm{CDT}}/k_{\mathrm{B}}$ & $130(5)\;\mu$K & $419(17)\;\mu$K  \\
   $N_{\mathrm{center}} ; N_{\mathrm{arms}} $ & $1.8(1)\times10^{6}$; $3.0(3)\times10^{6}$ & $6.4(4)\times 10^{4}$ ; $\sim 10^{5}$  \\
    $T$ & $24(1)\;\mu$K & $28(1)\;\mu$K\\
   $n_{\mathrm{peak}}$&$6.2(5)\times 10^{13}\;\mathrm{cm}^{-3}$ & $9.9(8)\times 10^{12}\mathrm{cm}^{-3}$ \\
PSD&  $1.4(2)\times 10^{-3}$ &$ 1.5(2)\times 10^{-3}$  
 \end{tabular}
 \end{ruledtabular}
 \end{table}
	
       \subsection{Evaporative cooling of a pure $^{161}$Dy cloud\label{Dyalone}}
             	
\begin{figure}
 \includegraphics[width=\columnwidth]{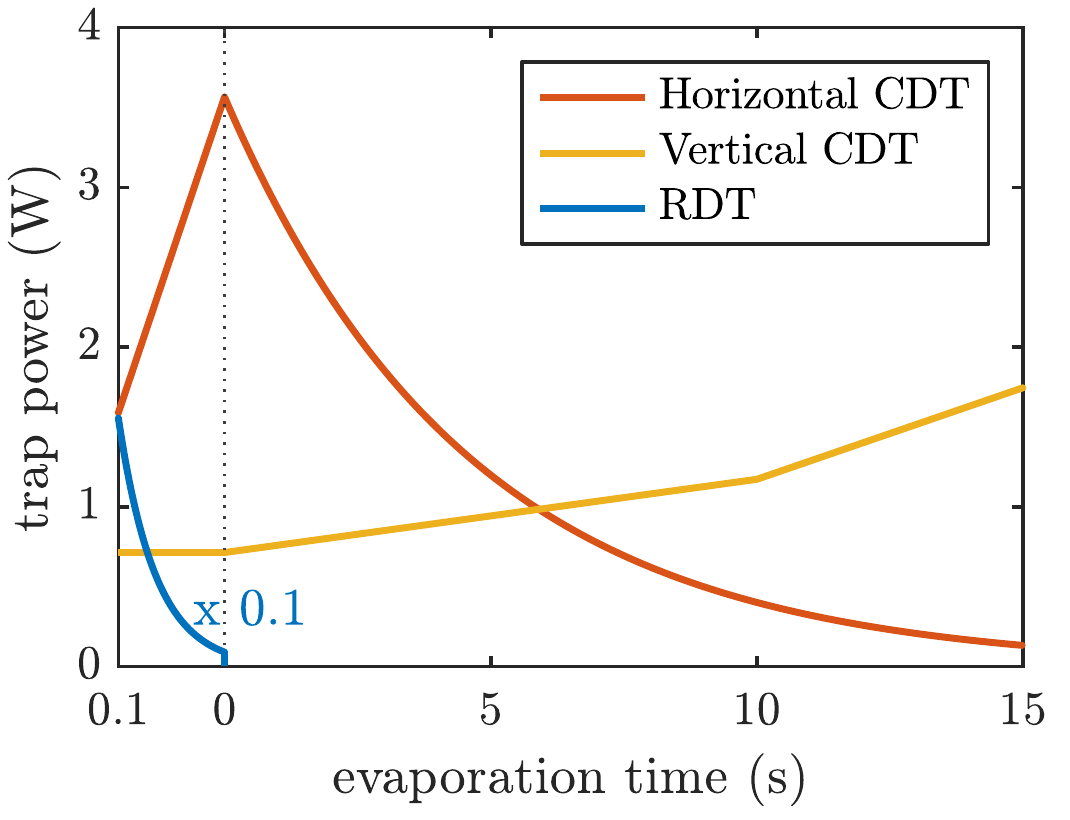}
 \caption{Evaporative cooling sequence. Power in the RDT (blue) and in the horizontal (red) and vertical (orange) beam of the CDT. The origin of time ($t=0$) corresponds to the start of the evaporation in the CDT. The power in the RDT and in the horizontal beam of the CDT are decreased in exponential ramps. The sequence is identical for the pure Dy cloud and for the Dy-K mixture. \label{evapsequence}}
 \end{figure}

       We now turn to the second phase of the evaporation, which is carried out in the CDT. We first consider the efficiency of the evaporative cooling of pure $^{161}$Dy, i.e.\ in the absence of K atoms. As soon as the RDT is switched off, the evaporation ramp is started. The magnetic field is kept at $430\;$mG, thus avoiding crossing any major Feshbach resonance~\cite{Baumann2014ool}. The evolution of the power in the two dipole traps is depicted in Fig.~\ref{evapsequence} and is optimized in terms of the final temperature ratio $T/T_{F}$ achieved, where $k_{\mathrm{B}}T_{F}=\hbar\bar{\omega}\left( 6N\right)^{1/3}$ is the Fermi temperature (in the following, we use the atom number in the central region of the CDT to calculate $T_{F}$). We adopt a standard evaporation scheme in which the power in the horizontal beam of the dipole trap is decreased while the power in the vertical one is increased~\cite{Takasu2003ssb,Stellmer2009bec,Lu2011sdb,Aikawa2012bec}. The power in the horizontal beam of the CDT is lowered in an exponential ramp to $150\;$mW within $15\;$s, while the power in the vertical beam is first increased linearly to $1.17\;$W within $10\;$s and then to $1.75\;$W within $5\;$s.
       
\begin{figure}
 \includegraphics[width=\columnwidth]{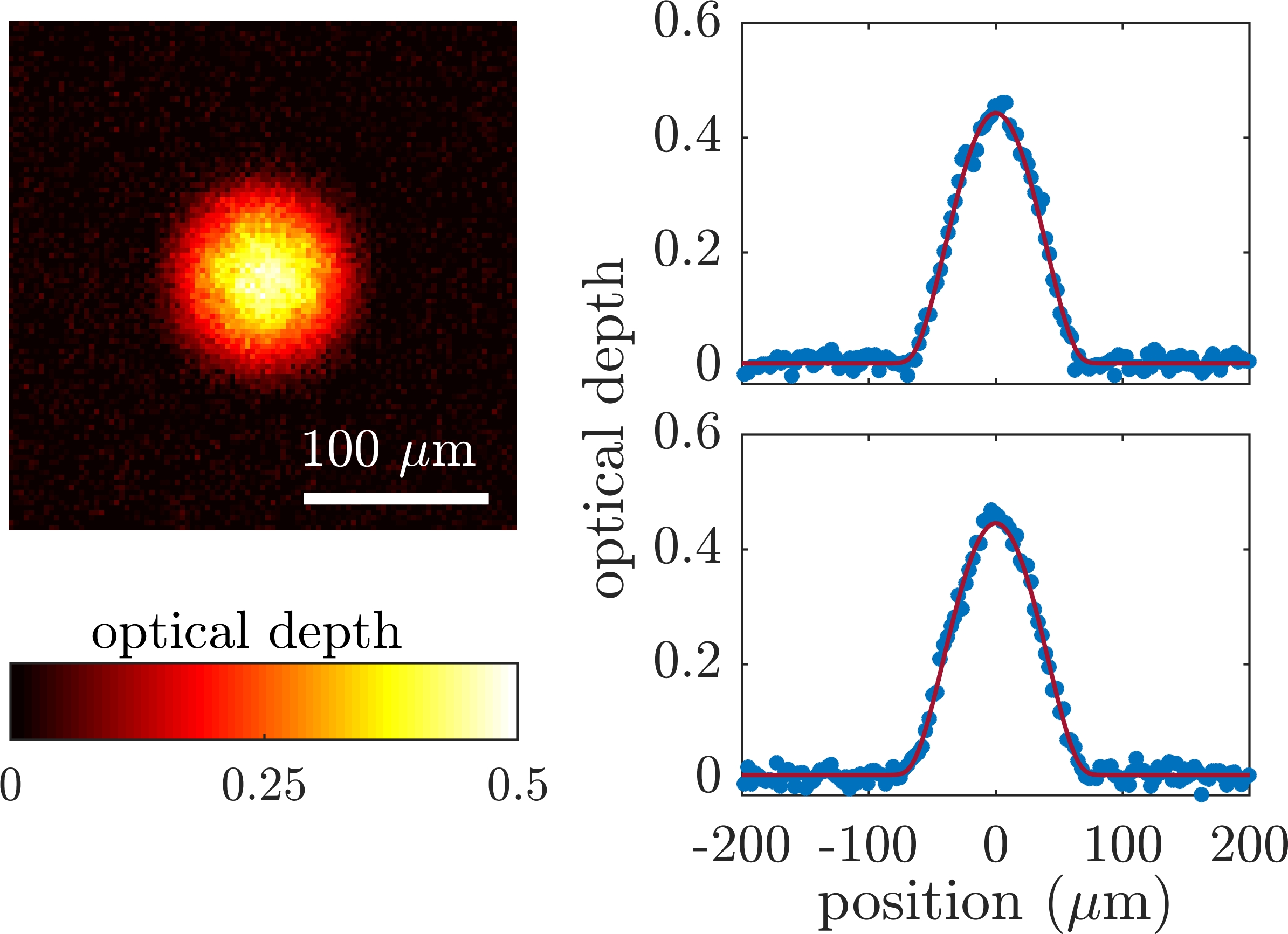}
 \caption{ Deep evaporative cooling of pure $^{161}$Dy. The density profile on the left-hand side is an average of ten absorption images of clouds of approximately $8\times 10^{4}$ $^{161}$Dy atoms taken after $14\;$s of evaporation in the CDT and a subsequent release from the trap (time of flight of $10\;$ms). The images on the right-hand side show cuts through the two-dimensional optical depth along the $x$- and $y$-axes. The solid lines show the corresponding profiles resulting from a two-dimensional polylogarithmic fit. From the fit we extract $T/T_{F}=0.085(10)$. \label{ODpic}}
 \end{figure}

        Figure~\ref{ODpic} shows a time-of-flight image of the degenerate cloud obtained at the end of the evaporation ramp (a brief description of our imaging scheme is presented in Appendix~\ref{imaging}). The two panels on the right show cuts along two orthogonal axes through the 2D density profile. The solid red lines result from a fit of the density profile by a polylogarithmic function. We obtain $8\times10^{4}$ atoms at a temperature of $60(10)\;$nK, corresponding to $T/T_{F}=0.085(10)$. The trap depth at this point is $U=k_{\mathrm{B}}\times720\;$nK and the trap frequencies are $[\omega_{x},\omega_{y}, \omega_{z}]=2\pi\times[133,248,176]\;$Hz, both calculated including the effect of gravity. The peak number density is $1.6\times 10^{14}\; \mathrm{cm}^{-3}$.
       
\begin{figure}
 \includegraphics[width=\columnwidth]{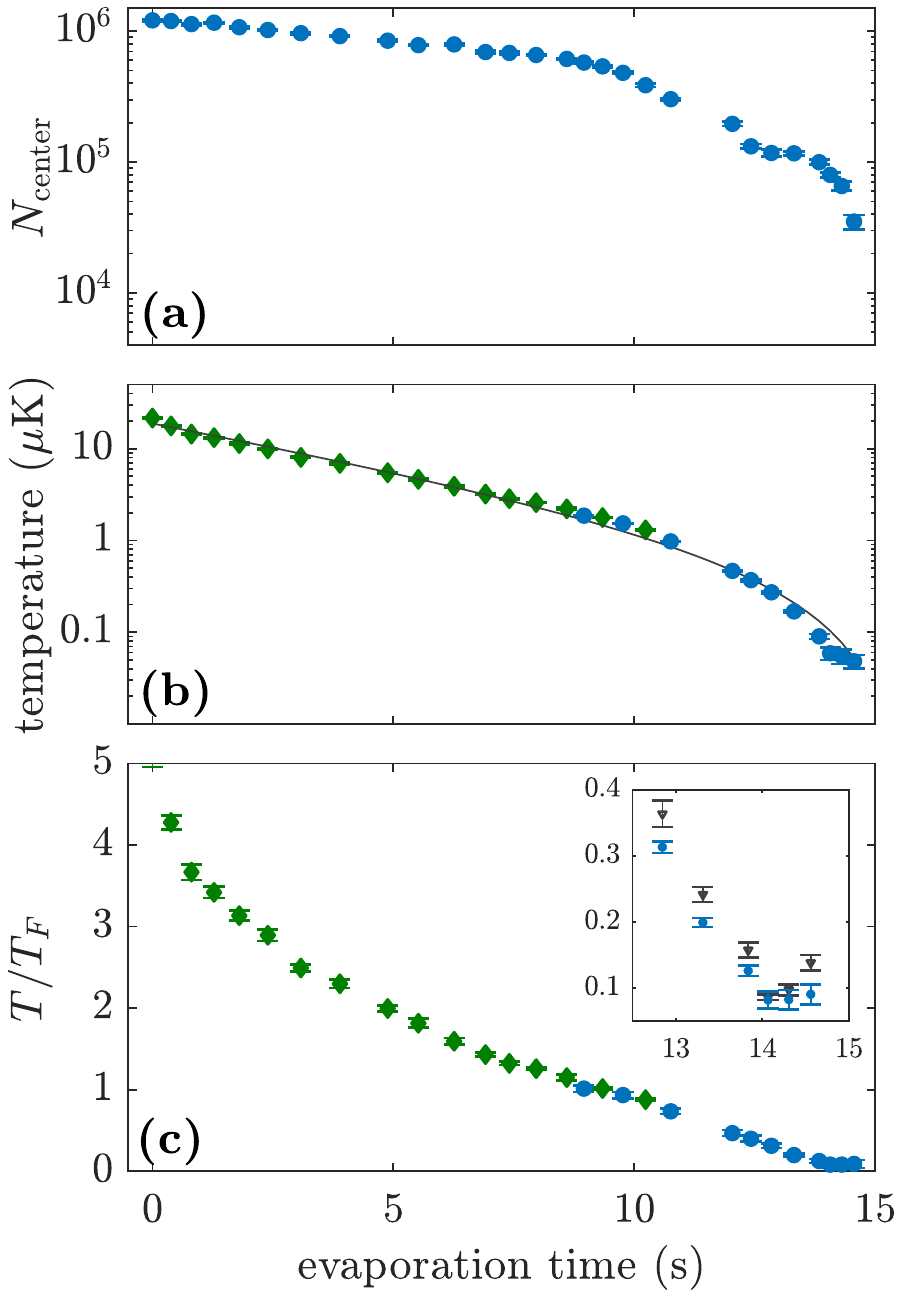}
 \caption{Evaporative cooling of $^{161}$Dy. Atom number, temperature and temperature ratio $T/T_{F}$ of the pure dysprosium cloud during the evaporation of the CDT.  In figures (b) and (c), red diamonds show the temperature extracted from a Gaussian fit of the column density, while blue points are extracted from a polylogarithmic fit. In (b), the solid black line shows the calculated trap depth reduced by a factor of seven. In (c) figure the inset shows a zoom-in on the final part of the evaporation ramp. Blue circles show the temperature ratio $T/T_{F}$ obtained from the size of the polylogarithm fit, while black triangles show the ratio obtained from the fitted fugacity through the relation $T/T_{F}=[-6\mathrm{Li}(\zeta)]^{-1/3}$. In all figures, the error bars show the standard error derived from ten repetitions. The time $t=0$ refers to the beginning of the evaporation in the CDT. \label{DyToTf}}
 \end{figure}

 In Figure \ref{DyToTf} we show the evolution of the atom number in the central region of the trap, of the temperature $T$ and of the ratio $T/T_{F}$ during the evaporation in the CDT. After approximately $10\;$s, we observe a change of slope in the evolution of the atom number in the central region, see Fig.~\ref{DyToTf}(a). We attribute this effect to the presence of atoms in the arms of the trapping potential. At $t=0$, around $3\times 10^{6}$  trapped Dy atoms are present in the horizontal arms. In the first $10\;$s, these atoms are then preferentially evaporated, cooling also the cloud in the central region. After $10\;$s, the arms of the trap are found to be empty and from that point on, the evaporation takes place in the central region of the potential only.  The initial fast gain in phase-space density (PSD $\propto \left(T/T_{F}\right)^{-3}$) associated with the slow decrease of the atom number in the central region suggests that the atoms in the arms play an important role in the efficiency of the forced evaporative cooling.
 
 The temperatures shown in Fig.~\ref{DyToTf}(b) are extracted from the observed column density profiles by Gaussian fits or by polylogarithmic fits. The first ones are applied at higher temperatures, the second ones at lower temperature. The fit by a polylogarithmic profile provides two parameters: the size $\sigma$, related to the temperature $T$, and the fugacity $\zeta$. The Fermi temperature is obtained from the measured atom number and the calculated averaged trap frequency. The fugacity on the other side gives direct access to $T/T_{F}=\left[-6\mathrm{Li}(-\zeta)\right]^{-1/3}$. The ratio $T/T_{F}$ shown in Fig.~\ref{DyToTf}(c) is obtained from the measured $T$ and the calculated $T_{F}$. A comparison with the second method using the fugacity is presented in the inset for the lowest temperatures achieved, and shows good agreement between the two methods. The change between the Gaussian and polylogarithmic fit models happens for $T/T_{F}\approx 1$. The ratio $T/T_{F}$ levels off after $14\;$s of evaporation. At this point, the trap depth is on the order of the Fermi energy $E_{F}=k_{\mathrm{B}}T_{F}$ and we enter the spilling regime, where efficient evaporation stops. 
 
In Fig.~\ref{DyToTf}(b) we compare the measured temperature $T$ with the calculated trap depth $U$ (where the effect of gravity is taken into account). We find that the forced evaporation is well characterized by a truncation parameter $\eta=U/k_{\mathrm{B}}T\approx 7$ throughout the whole evaporation process. 

The trapping potential created by the CDT is deformed by gravity. In particular, the trap depth is reduced and, in the very final stage of the evaporation, the trap frequency along the vertical axis is lowered. In our scheme, after $14\;$s of evaporation, the depth of the CDT is reduced from $5.3\;\mu$K (depth without gravity) to $720\;$nK, and the trap frequency along the vertical axis is reduced from $210\;$Hz to $176\;$Hz. The reduction of the trap depth is actually beneficial for the evaporative cooling~\cite{Hung2008aec}. Indeed, the reduction achieved by decreasing the laser power is inherently accompanied by a lowering of the trapping frequencies, which in turn implies a lower collision rate and a less efficient evaporative cooling. Gravity reduces the trap depth essentially without affecting the confinement strength of the trap (the lowering of the trap frequencies only appears at very low trap depth), and allows us to maintain a higher collision rate at low trap depths.

We achieved the lowest temperature $T/T_{F}=0.085(10)$ after $14\;$s of evaporation. For longer evaporation times the decreasing trap depth reaches the Fermi energy and we observe spilling of the dysprosium cloud. As in earlier work on evaporative cooling of fermions (see e.g.~\cite{Lous2017toa}), we find that the optimal cooling is achieved just before the onset of spilling.
               
        \subsection{Evaporative cooling of the $^{161}\mathrm{Dy}$-$^{40}\mathrm{K}$ mixture\label{DyandK}}
              
         We now characterize the combined cooling of the Fermi-Fermi mixture. Starting with both components loaded into the RDT, the experimental procedure is identical to the one described for Dy alone, as we find that it also provides the best results for the mixture, provided that the K atom number is kept small (typically, MOT loading time shorter than $3\;$s, resulting in a final atom number of a few $10^{3}$). Indeed we observe that an increase of the K atom number (achieved by extending the K MOT loading) compromises the cooling of the Dy sample. As before, the strength of the magnetic field is set to $430\;$mG.
                  
\begin{figure}
 \includegraphics[width=\columnwidth]{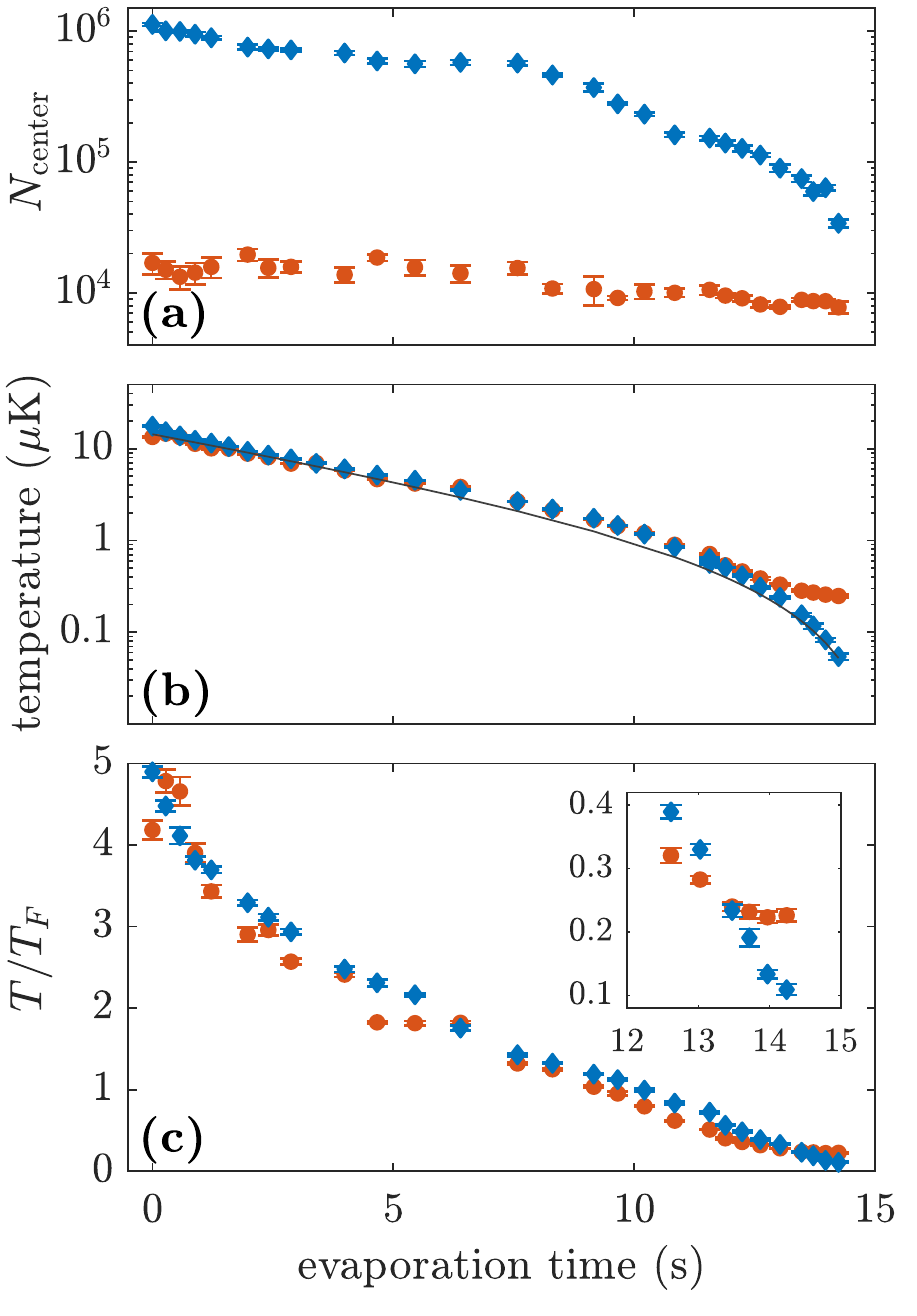}
 \caption{Sympathetic cooling of $^{40}$K by the evaporatively cooled $^{161}$Dy. Atom number, temperature and temperature ratio $T/T_{F}$ of the $^{40}$K (red circles) and $^{161}$Dy (blue diamonds) samples during the evaporation of the CDT. As in Figure~\ref{DyToTf}, the solid black line in (b) shows the trap depth for Dy reduced by a factor $7$. The error bars show the standard error derived from ten repetitions. As in Figure \ref{DyToTf}, the time $t=0$ refers to the beginning of the final ramp of the CDT. \label{DyKevap}}
 \end{figure}

         We first describe the process of evaporative cooling. The atom number in the central region of the trap and the temperature of the two samples in the CDT are shown in Fig.~\ref{DyKevap}. In these measurements, the atoms are held at fixed power during $10\;$ms before a time of flight expansion is performed. The behavior of the number of Dy atoms is essentially the same as observed in the evaporation of pure Dy, with the same change of slope after $10\;$s of evaporation. The number of K atoms on the other hand decreases very slowly, its final value being approximately half the initial one. Given the deep potential seen by the K atoms, we can exclude evaporation losses, and we therefore attribute the decrease in K atom number to inelastic processes. 
         
         The equality of the two temperatures seen in Fig.~\ref{DyKevap}(b) demonstrates the efficiency of the sympathetic cooling. However, we observe that the temperature of the K sample decouples after approximately $13\;$s in the evaporation ramp from the temperature of the Dy component. While the temperature of the Dy follows the evolution of the trap depth, the K temperature decreases much more slowly. As a consequence of this behavior, the ratio $T_{\mathrm{K}}/T_{F,\mathrm{K}}$ essentially levels off, at a value close to 0.2. This observed decoupling is not an effect of the gravitational sag~\cite{Hansen2013poq}, as we verified by applying a levitation field (see Sec.~\ref{levitation}). We speculate this is an effect of Pauli blocking in the deeply degenerate Dy cloud, which strongly reduces the rate of elastic collisions between the species. Further investigation is needed to understand the final limitations of this cooling scheme. 
         
\begin{figure}
 \includegraphics[width=\columnwidth]{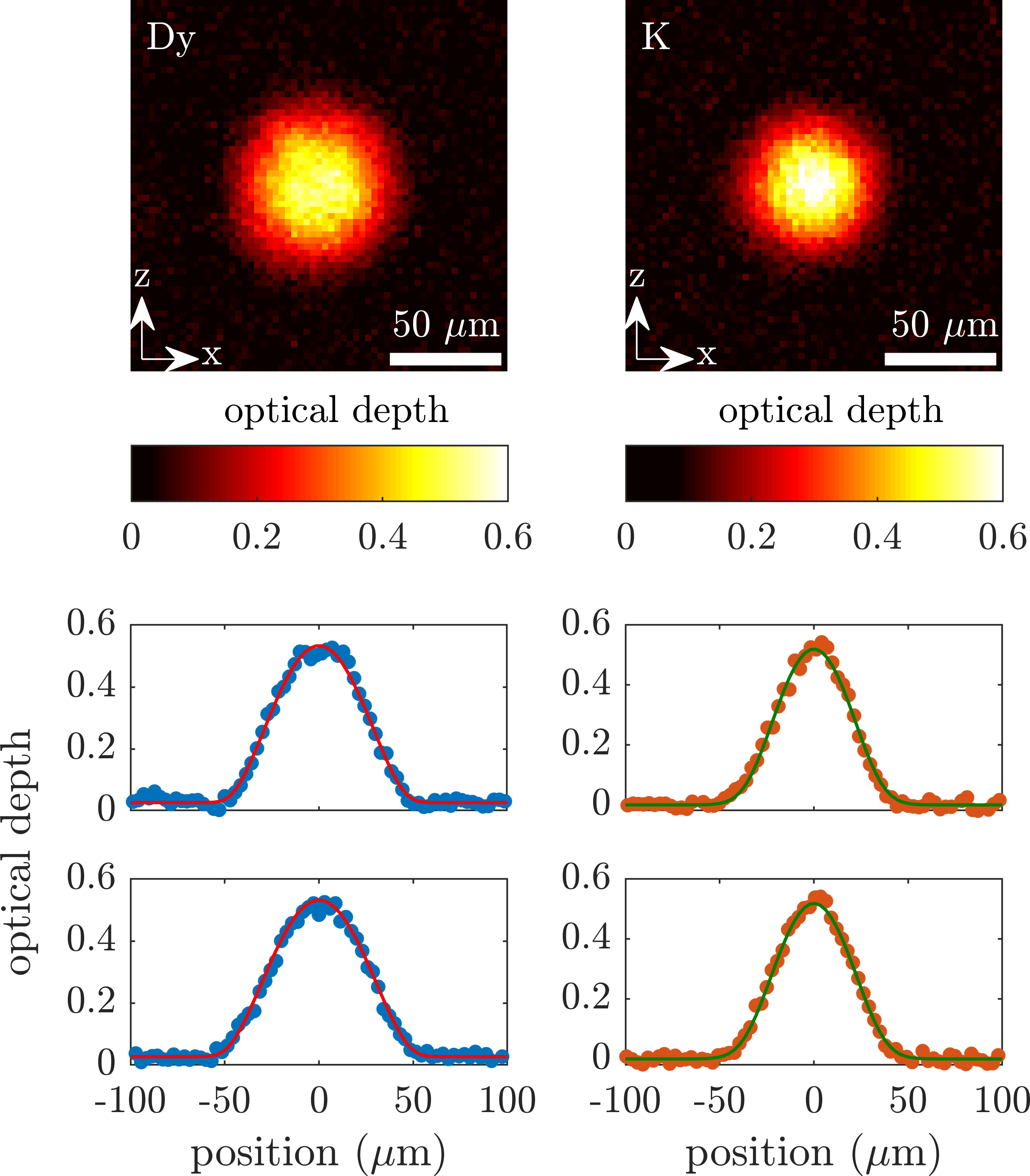}
 \caption{Deep cooling of the $^{161}$Dy-$^{40}$K mixture. Density profile of a cloud of $^{161}$Dy ($^{40}$K ) atoms after a time of flight expansion of $7.5\;$ms ($2.5\;$ms) for a ratio $T_{\mathrm{Dy}}/T_{F,\mathrm{Dy}}=0.09(1)$ ($T_{\mathrm{K}}/T_{F,\mathrm{K}}=0.22(2)$). Atom numbers are $N_{\mathrm{K}}=9(1)\times10^{3} $ and $N_{\mathrm{Dy}}=4.1(4)\times10^{4}$. The upper panel shows the averaged column density of six images obtained with resonant absorption imaging. The lower panels show cuts of the column density along the $x$ (top) and $z$ (bottom) axes. The solid lines are cuts of a 2D fit with a polylogarithmic function.\label{doubleOD}}
 \end{figure}

          Figure~\ref{doubleOD} shows our final cooling results for the evaporation ramped discussed before. The absorption images were taken after $14.2\;$s of evaporation and a time of flight expansion of $2.5\;$ms for K and $7.5\;$ms for Dy. The lower panel shows cuts through the 2D profile, and the solid lines show the corresponding fits by a polylogarithmic function. We obtain $9\times 10^{3}$ K atoms and $4.1\times 10^{4}$ Dy atoms. At this point the two Fermi energies are calculated to be $E_{F,\mathrm{K}}/k_{B}=1.13\;\mu$K for K and $E_{F,\mathrm{Dy}}/k_{B}=500\;$nK for Dy. We measure temperatures of $T_{\mathrm{K}}=250(10)\;$nK and $T_{\mathrm{Dy}}=45(4)\;$nK, respectively, corresponding to $T/T_{F}=0.22(2)$ for K and $T/T_{F}=0.09(1)$ for Dy. 
         
          By varying the K MOT loading time we can control the number of $^{40}$K atoms in the RDT and the resulting population ratio of the two species at the end of the evaporation. Indeed, after the transfer in the RDT, the K cloud is hotter than the Dy one and a larger number of K atoms thus represents a larger heat load for the Dy cloud, affecting its evaporation dynamics. For final number ratios $N_{\mathrm{Dy}}/N_{\mathrm{K}}$ from approximately 4 (corresponding to the situation described in Figs.~\ref{DyKevap} and~\ref{doubleOD}) down to 2, we obtain similar cooling performance, with the two species as cold as $T/T_{F}\approx 0.2$. Then, if we further reduce the number ratio by increasing the initial number of K atoms, the K component reaches lower $T_{\mathrm{K}}/T_{F,\mathrm{K}}$ ratios at the expense of $T_{\mathrm{Dy}}/T_{F,\mathrm{Dy}}$. The absolute temperatures $T_{\mathrm{Dy}}$ and $T_{\mathrm{K}}$ achieved actually only weakly depend on the population imbalance. The main effect of the number ratio on the respective $T/T_{F}$ comes from the dependence of the Fermi temperature on the atom number. In the extreme case where the Dy atom number is zero at the end of the evaporation, we achieve $T/T_{F}\approx 0.15$ for K. 
           
           %The low initial atom number is a consequence of the transfer from the RDT into the CDT. K atoms are efficiently transferred in the two beams of the CDT, but a large fraction of the atoms trapped in the wings of the potential does not accumulate in the central region and is lost when the power of the CDT is lowered.
           
       \subsection{Effect of magnetic levitation\label{levitation}}
       
      Mixing two species in an optical dipole potential requires careful consideration of the effect of gravity~\cite{Hansen2013poq}. The effect on the total potential seen by each species is twofold: Its minimum is shifted in the vertical direction (referred to as gravitational sag), and its depth is reduced. Because of the different masses and polarizabilities, these two effects have different magnitudes for the two species. In our mixture, gravity has a much stronger effect on Dy than on K, because of both the larger mass and the weaker polarizability. As discussed in Sec.~\ref{Dyalone}, when we reach optimum cooling results for Dy after 14s of evaporation, the potential depth is reduced by gravity to only $14\%$ of the optical potential depth. In contrast, the K trap depth is only reduced to $88\%$ of the optical potential depth. At the same point the potential seen by the Dy atoms is shifted by approximately 7$\;\mu$m, which is already comparable to the size of the two clouds, while the sag experienced by the K atoms is small (around $500\;$nm). 
      
      When considering a single paramagnetic species, the common way to cancel the effect of gravity is to apply a gradient of magnetic field such that the spatially varying Zeeman energy creates a force opposing the gravitational force~\cite{Weber2003bec,Anderson1995oob}. The atomic cloud is then levitated. Because of their different magnetic dipole moments and masses, the strength of the gradient of magnetic field for levitation is different for two different atomic species. Yet, there exists a ``magic'' value of the gradient such that the gravitational sag of the two species is identical for any power of the optical dipole trap~\cite{Lous2017toa}. In the case of Dy and K, the much stronger dipole moment of Dy atoms brings this ``magic'' value ($2.69\;$G/cm) very close to the value for the levitation of dysprosium ($2.83\;$G/cm). 
      
  \begin{figure}
 \includegraphics[width=\columnwidth]{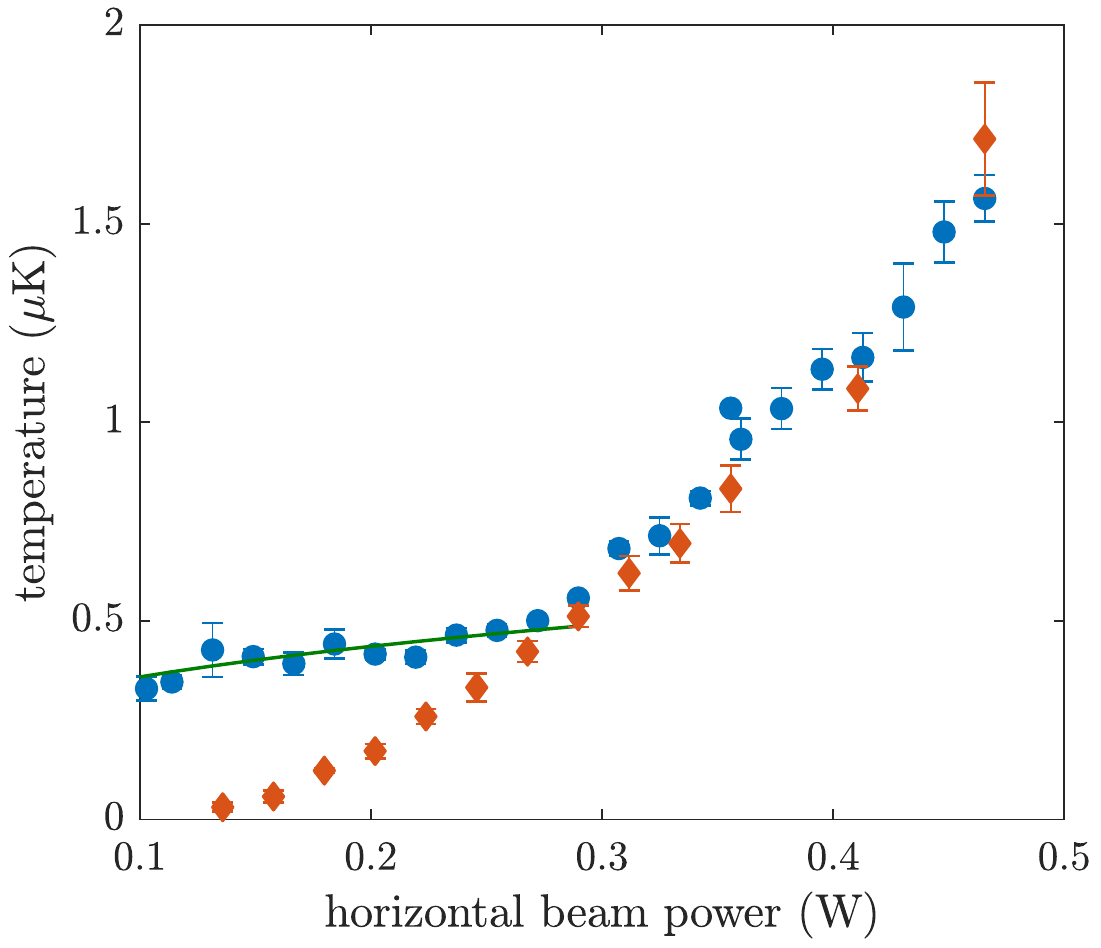}
 \caption{Effect of levitation on evaporative cooling. The temperature of a cloud of Dy is measured during the evaporation of the CDT in the absence of levitation (red diamonds) and when the levitation fields are switched on below $0.3\;$W (blue points). The solid green line is a fit of the temperature below $0.30\;$W by an adiabatic cooling model assuming that the temperature for a power $P_{f}$ in the trap is proportional to the average trap frequency $\bar{\omega}$ at this power: $T(P_{f})=b\; \bar{\omega}(P_{f})$ (see main text). The error bars show the standard error derived from six repetitions. \label{levitationfig}}
 \end{figure}  
           
      We demonstrate the effect of the levitation gradient by performing the following experiment. We evaporate dysprosium atoms in the CDT down to a given final power of the horizontal beam and turn on a levitation field when the ramp of power reaches below a threshold value set to $0.3\;$W. We then hold the cloud in the trap for $100\;$ms and finally release it for $8\;$ms. We extract the temperature from a polylogarithmic fit to the two-dimensional density profile. The measured temperatures are presented in Fig.~\ref{levitationfig}. We observe that the evaporation essentially stops when the levitation is turned on, which is a consequence of the corresponding increase of the trap depth. With decreasing power, the temperature still decreases slowly. The green line in Fig.~\ref{levitationfig} is a fit of the temperature for final powers lower than $0.3\;$W by a function of the form $T(P_{f})=b\; \bar{\omega}(P_{f})$, where $\bar{\omega}(P_{f})$ is the calculated average trap frequency for a final trap power $P_{f}$, and $b$ is a fit parameter. The good agreement between the fit and the measured temperatures shows that the decrease of the temperature in the presence of levitation can be fully attributed to the adiabatic opening of the confinement. 
      
Levitation is an efficient technique to ensure the spatial overlap of two species, and will be a very useful tool in further experiments investigating the physics of Fermi-Fermi mixtures, where full overlap of the two constituents is required. Yet, in our scheme, even in the absence of levitation the two species remain in thermal contact throughout the evaporation ramp, as shown by the efficiency of the sympathetic cooling. This means that we can  apply the levitation after the evaporation ramp and benefit from both the positive effects of gravity on evaporative cooling and of levitation on the spatial overlap.          

       \subsection{Interspecies scattering cross section\label{crossedthermalization}}
       
	To determine the cross section for elastic collisions between $^{161}$Dy and $^{40}$K atoms we perform an interspecies thermalization measurement~\cite{Mosk2001mou,Tassy2010sci,Guttridge2017iti}. We interrupt the evaporation process in the CDT when the power in the horizontal beam reaches 350$\;$mW. At this point, there are no atoms left in the arms of the CDT. Then we recompress the trap by increasing the power in the horizontal beam to 680$\;$mW to stop plain evaporation. The two atom numbers are $N_{\mathrm{K}}\approx 2\times 10^{4}$ and $N_{\mathrm{Dy}}\approx 10^{5}$. The two components are in the thermal regime and their temperatures are equal, around $2\;\mu$K. The average trap frequency is $\bar{\omega}= 2\pi\times1007\;$Hz for K atoms, and the peak densities are $n_{\mathrm{K}}= 3.6\times 10^{13}\;\mathrm{cm}^{-3}$ and $n_{\mathrm{Dy}}=3.1\times 10^{13}\;\mathrm{cm}^{-3}$. We then suddenly displace the horizontal beam in the vertical direction using an AOM and, after a time $\delta t$, suddenly displace it back to its original position.  After this sequence of two trap displacements, we hold the atoms in the CDT for a variable time and finally measure the temperature of the two components. By setting $\delta t=2\pi/\omega_{z}^{\mathrm{Dy}}$ the energy injected in the Dy cloud by the two successive kicks ideally vanishes, while it does not for the K cloud. This kinetic energy of the K atoms is transformed after some typical time $\tau$ in thermal energy through collisions with Dy atoms. We measure this thermal energy after a variable hold time by measuring the size of the K cloud after a time of flight expansion. Following the model developed in \cite{Mosk2001mou} and considering the zero energy limit for elastic collisions, the total collision rate is obtained from $\Gamma_{\mathrm{coll}}=\sigma_{\mathrm{el}} v_{\mathrm{rel}}I$, where $\sigma_{\mathrm{el}}$ is the scattering cross section for elastic collisions between Dy and K atoms, $v_{\mathrm{rel}}$ is the mean relative velocity of two colliding atoms and $I=\int{n_{\mathrm{K}}n_{\mathrm{Dy}}\mathrm{d}V}$. The relation between the measured thermalization time $\tau$ and the total collision rate is described in Appendix \ref{AppendThermalization}. 
		
	Figure \ref{thermalization} shows a thermalization curve, where the measured size of the K sample is interpreted in terms of a temperature~\footnote{Since the $^{40}$K cloud cannot thermalize by itself after the excitation, interspecies collisions are needed to reach thermal equilibrium. For shorter times, the $^{40}$K cloud can therefore not be described by a temperature in a strict thermodynamic sense. Here we use an effective temperature as a measure of the cloud's kinetic energy in the center-of-mass frame.}. The temperatures of the two species approach each other as we hold the cloud in the trap. Because of the number ratio $N_{\mathrm{Dy}}/ N_{\mathrm{K}}\approx 5$, the dominant effect is observed as a reduction of the K temperature, while the Dy temperature only slightly increases. Our observation is consistent with a simple heat transfer from the K sample to the Dy one. Atom numbers of both species are constant on the time scale of the measurement.

	The temperatures of the two species enter in the value of $v_{\mathrm{rel}}$ and of the overlap integral $I$; see Appendix~\ref{AppendThermalization}. It turns out that, given the mass and polarizability ratios of Dy and K, the opposite effects of temperatures on $v_{\mathrm{rel}}$ and $I$ nearly balance each other and thus lead to an almost constant value of $\Gamma_{\mathrm{coll}}$ for a temperature ratio $T_{\mathrm{K}}/T_{\mathrm{Dy}}\lesssim 2$. Since at these temperatures the elastic scattering cross section is almost constant, the measured temperatures evolve in a quasi-exponential way during the thermalization process. From an exponential fit of the temperature of K for hold times larger than $80\;$ms we extract a thermalization time constant of $\tau=130\;$ms. This corresponds to an elastic scattering cross section of $\sigma_{\mathrm{el}}\approx 5.9\times 10^{-17}\;\mathrm{m}^{2}$. We calculate the cross section for elastic dipolar collisions between Dy and K atoms to be two orders of magnitude smaller than the measured cross section~\footnote{The elastic dipolar cross section for interspecies collisions is obtained from $\sigma_{\mathrm{dip}}=(16/45)\pi a_{D}^{2}+(16/15)\pi a_{D}^{2}$, where the two terms represent the contributions of even and odd partial waves respectively. The dipolar length $a_{D}$ is given by $a_{D}=\mu_{0}\mu_{\mathrm{K}}\mu_{\mathrm{Dy}}m_{r}/(4\pi \hbar^{2})$. Here $\mu_{0}$ is the permeability of vacuum, $\mu_{\mathrm{K}}$ and $\mu_{\mathrm{Dy}}$ are the magnetic dipole moments of K and Dy atoms in the relevant spin states, and $m_{r}$ is the reduced mass. We calculate $\sigma_{\mathrm{dip}}=7.2\times 10^{-19}\mathrm{m}^{2}$}. We therefore attribute the measured cross section to the contact interaction and deduce the corresponding scattering length $\vert a_{\mathrm{DyK}}\vert \approx 40\;\mathrm{a}_{0}$, with an estimated error of $30\%$ including all statistical and systematic errors.
	
\begin{figure}
 \includegraphics[width=\columnwidth]{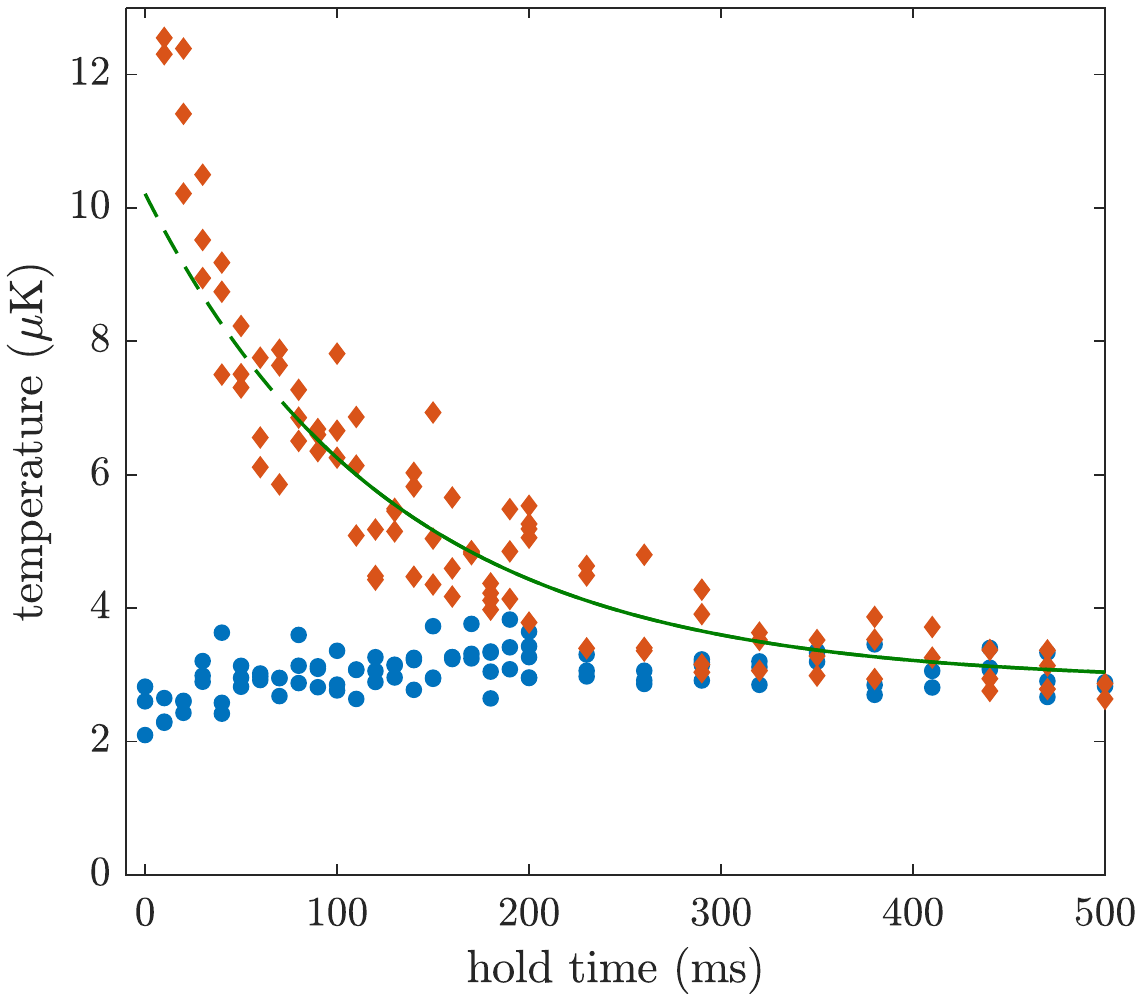}
 \caption{Cross-species thermalization measurement. The temperatures of the $^{40}$K (red diamonds) and $^{161}$Dy (blue circles) are shown versus the hold time after the species-selective heating by the trap displacements (see text). The solid green line is a fit of the temperature of the K atoms by an exponential curve for a hold time bigger than $80\;$ms. The dashed line extrapolates the fit.  \label{thermalization}}
 \end{figure}

\section{Conclusion and Outlook\label{conclusion}}
%Give brief summary and discuss main conclusions from our work .
We have demonstrated how a deeply degenerate Fermi-Fermi mixture of ultracold $^{161}$Dy and $^{40}$K atoms can be efficiently produced in an optical dipole trap. The cooling process relies on the evaporation of the spin-polarized Dy component, with elastic collisions resulting from universal dipolar scattering. The K component is sympathetically cooled by elastic $s$-wave collisions with Dy. In this way, we have reached conditions with about $4\times 10^{4}$ Dy atoms and about $10^{4}$ K atoms at temperatures corresponding to $\sim$10\% and $\sim$20\% of the respective Fermi temperature. This represents an excellent starting point for future experiments aiming at the realization of novel quantum phases and superfluid states in mass-imbalanced fermionic mixtures. 

The next challenge in our experiments will be the implementation of interspecies interaction tuning by means of magnetically tuned Feshbach resonances \cite{Chin2010fri}. Although the particular interaction properties of our mixture are yet unknown, one can expect many Feshbach resonances to arise from the anisotropic electronic structure of the Dy atom \cite{Petrov2012aif, Kotochigova2014cib}. The key question, which we will have to answer in near-future experiments, is whether sufficiently broad Feshbach resonances exist to facilitate interaction tuning along with a suppression of inelastic losses. Reason for optimism is given by the fact that rather broad Feshbach resonances have been found in single-species experiments with Er and Dy \cite{Frisch2014qci, Maier2015buf, Lucioni2018ddb}.
%This will be a key issue for the realization of collisionally stable many-body states under conditions of strong interactions.

If nature will be kind to us and provide us with a good handle to control interactions in the resonant regime, then the future will look very bright for novel superfluid phases in Dy-K Fermi-Fermi mixtures. A very favorable property of this mixture, as we pointed out already in Ref.~\cite{Ravensbergen2018ado}, is the fact that the polarizability ratio of the two species in an infrared optical dipole trap nearly corresponds to the inverse mass ratio. This will allow to easily match the Fermi surfaces of both species even in the inhomogeneous environment of a harmonic trap, and to investigate pairing and superfluidity of unequal mass particles in the crossover from molecular BEC to a BCS-type regime. Moreover, our preparation naturally produces a situation with a majority of heavier atoms, which is exactly what will be needed to realize FFLO states in mass- and population-imbalanced fermionic mixtures \cite{Gubbels2013ifg, Wang2017eeo}.
% Surround figure environment with turnpage environment for landscape
% figure
% \begin{figure}
% \end{figure}

% tables should appear as floats within the text
% The ruledtabular enviroment adds doubled rules to table and sets a
% reasonable default table settings.
% Use the table* environment to get a full-width table in two-column
% Add \usepackage{longtable} and the longtable (or longtable*}
% environment for nicely formatted long tables. Or use the the [H]
% placement option to break a long table (with less control than 
% \begin{table}%[H] add [H] placement to break table across pages
% \begin{ruledtabular}
% Lines of table here ending with \\
% \end{tabular}
% \end{ruledtabular}
% \end{table}

% Surround table environment with turnpage environment for landscape
% table
% \begin{table}
% \begin{ruledtabular}
% \end{tabular}
% \end{ruledtabular}
% \end{table}

\begin{acknowledgments}
We thank Slava Tzanova for her contributions in the early stage of the experiment. We acknowledge support by the Austrian Science Fund (FWF) within the Doktoratskolleg ALM (W1259-N27). 
\end{acknowledgments}

\appendix

\section{Apparatus\label{apparatus}}

Here we describe briefly our apparatus. We use two independent and spatially separated atomic sources for the two species, which provide collimated atomic beams in the main vacuum chamber through two different ports. As mentionned in Sec.~\ref{MOT}, our source of Dy atoms consists of a commercial high-temperature effusion oven operating at about $1000^{\circ}$C combined with a Zeeman slower. The atomic flux through the Zeeman slower is increased by a transverse cooling stage applied at the exit of the oven, operating on the same broad transition as the Zeeman slower. 

The K source is a two-dimensional MOT in a glass cell~\cite{Dieckmann1998tdm}, loaded from the background pressure created by commercial isotopically enriched K dispensers (Alvatec), placed directly inside the glass cell. The glass cell is connected to the main vacuum chamber by a differential pumping tube. The two-dimensional MOT operates on the D2 line of K, and is created by two perpendicular retro-reflected elliptical laser beams and by a two-dimensional quadrupole magnetic field realized by two pairs of coils. The transversally cooled atoms are trapped around the line of zero magnetic field, forming a cigar-shaped cloud. An additional beam pushes the cloud through the differential pumping tube in the main chamber.

The K and Dy three-dimensional MOT beams are overlapped in a four fiber cluster, which distributes the light to the four MOT beam paths. The MOTs are created by two retro-reflected orthogonal beams in the horizontal plane, and by two counter-propagating vertical beams. The narrow character of the $626\;$nm line causes the Dy cloud to sit below the zero of the magnetic field, hence requiring the large diameter MOT beams ($35\;$mm in our case). The laser light at 626~nm is spectrally broadened by a high-efficiency electro-optical modulator resonant at 104~kHz, which adds around 30 sidebands. 

\section{Imaging of the atomic cloud\label{imaging}}

The K and Dy atoms are imaged through the same imaging set-up, on a single sCMOS camera (Andor Neo), along an axis that lies in the horizontal plane (about $40^{\circ}$ to the $x$-axis). We use resonant absorption imaging on the D2 line of wavelength $767\;$nm for K, and on the broad line at $421\;$nm for Dy. The two probe beams are overlapped on a dichroic mirror and have $\sigma^{-}$ polarization. The imaging sequence involves four laser pulses and one background picture: each species requires one pulse to image the atoms followed by one normalization pulse. The background picture without probe light is taken in the end of the imaging sequence. From the moment where the CDT is switched off, we wait a time corresponding to the time of flight for K,  and shine the pulse imaging K atoms for $10\:\mu$s. We then wait a variable time which sets the difference between the two values for the time of flight applied to K and Dy, and shine the pulse imaging Dy atoms for $10\;\mu$s. The minimum time between the two pulses is $500\;$ns. The two normalization pictures are taken $40\;$ms after the respective pictures of the atoms. In both cases, the scattering cross section is assumed to be the resonant cross section for a two-level system.

\section{Cooling $^{40}$K on the D1 line\label{D1cooling}}

After preparation in the MOTs, the K and Dy components have very different temperatures. However, optimal efficiency of the subsequent evaporative cooling stage requires similar starting conditions for the two species. In order to reach this situation, we perform after the MOT stage sub-Doppler cooling of the K atoms in the form of a gray molasses on the D1 line. While this cooling technique has been demonstrated in free space~\cite{fernandes2012sdl}, we describe here its successful implementation in the presence of our large volume reservoir optical dipole trap (Sec.~\ref{sequence}). In this situation, the dissipative nature of the molasses cooling improves significantly the phase-space density reached after transfer into the optical dipole trap. We account for the effect of the space-dependent light shift by ramping the laser parameters, as we describe in the following.

\begin{figure}
 \includegraphics[width=\columnwidth]{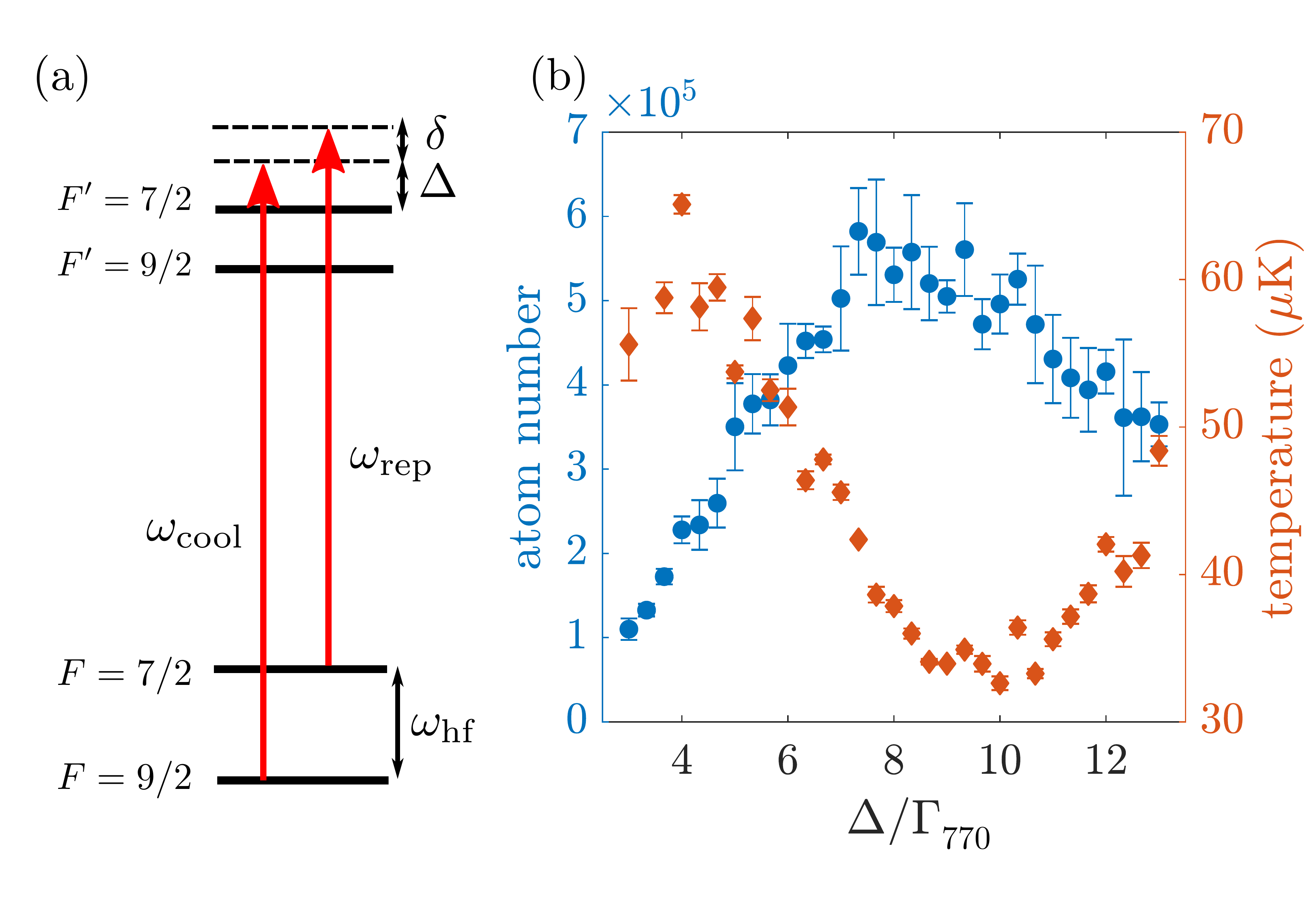}
 \caption{D1 cooling of $^{40}$K. (a) levels structure of the D1 transition and laser scheme. (b) Atom number and temperature at the end of the D1 cooling step in the RDT, as a function of the detuning $\Delta$, when $\delta$ is kept to zero.\label{D1levels}}
 \end{figure}

The laser beams creating the gray molasses are overlapped on polarizing beam splitters with the MOT beams, which fixes their polarization to circular with opposite helicity compared to the MOT beams. The cooler and repumper beams are blue-detuned relative to the $F=9/2\rightarrow F'=7/2$ and $F=7/2\rightarrow F'=7/2$ transitions respectively. We define $\Delta$ as the detuning of the cooling beam relative to the bare (in the absence of light shift) $9/2\rightarrow 7/2$ line and $\delta=\omega_{\mathrm{rep}}-\omega_{\mathrm{cool}}+\omega_{\mathrm{hf}}$, where $\omega_{\mathrm{cool}}$ is the frequency of the cooling beam, $\omega_{\mathrm{rep}}$ is the frequency of the repumper and $\omega_{\mathrm{hf}}$ is the hyperfine splitting in the ground state of $^{40}$K; see Fig.~\ref{D1levels}(a). 

After the compression of the K MOT, the gradient of magnetic field is set to zero, and residual homogeneous magnetic fields are canceled to the level of $10\;$mG. The D2 light is switched off and D1 light is switched on. The RDT is turned on beforehand during the MOT and its power is kept unchanged (see Sec.~\ref{sequence}). At this point the cloud sits on top of the crossed RDT, though being larger. 

Because the polarizability of the excited state at $1064\;$nm is negative, the transition is shifted to higher frequency by the AC Stark shift in the center of the trap compared to the wings of the potential. We therefore use a two-step D1 cooling scheme, in which the intensities and the detuning $\Delta$ are first kept constant and then tuned in linear ramps. During the initial capture phase of $0.6\;$ms, the power and detunings are kept constant at $I^{\mathrm{cool}}=5.8\;I_{\mathrm{sat},770}, I^{\mathrm{rep}}=0.11\;I_{\mathrm{sat},770}$ and $\delta=0,  \Delta=3\;\Gamma_{770}$ (where $I_{\mathrm{sat},770}= 1.70\;\mathrm{mW\ cm}^{-2}$ is the saturation intensity of the D1 line of K and $\Gamma_{770}$ its natural linewidth), which are the parameters found to be optimal for D1 cooling in free space. In the second phase, powers are ramped down in $20\;$ms to $I^{\mathrm{cool}}=0.18\;I_{\mathrm{sat}}$ and $I^{\mathrm{rep}}\approx0.01\;I_{\mathrm{sat}}$ while the cooler and repumper are detuned further to the blue. The ramp of frequencies thus allows us to cool atoms from the outer region into the trap. Figure~\ref{D1levels}(b) shows the final atom number and temperature of the K cloud as the final value of $\Delta$ is varied. We achieve a temperature of $27(1)\;\mu$K for a final value $\Delta=9\;\Gamma_{770}$.

\section{Interspecies thermalization\label{AppendThermalization}}

We summarize here the model developed by Mosk et al.~\cite{Mosk2001mou}, which describes the thermalization between two species in a harmonic trap (we assume here that the two species have the same central position). We reproduce in particular the relation between the elastic cross section $\sigma_{\mathrm{el}}$ and the thermalization time $\tau$. First we calculate the total collision rate $\Gamma_{\mathrm{coll}}=\sigma_{\mathrm{el}} v_{\mathrm{rel}}I$. The mean relative velocity is obtained as
\begin{equation}
 v_{\mathrm{rel}}=\sqrt{\frac{8k_{\mathrm{B}}}{\pi}\left(\frac{T_{\mathrm{K}}}{m_{\mathrm{K}}}+\frac{T_{\mathrm{Dy}}}{m_{\mathrm{Dy}}}\right)}\ .
 \label{equa1}
 \end{equation}
 Assuming a thermal density distribution of the two components, one calculates the overlap integral 
\begin{equation} 
I=N_{\mathrm{K}}N_{\mathrm{Dy}}\bar{\omega}_{\mathrm{Dy}}^{3}\left(\frac{2\pi k_{\mathrm{B}}T_{\mathrm{Dy}}}{m_{\mathrm{Dy}}}\right)^{-3/2}\left(1+\frac{\alpha_{\mathrm{Dy}}}{\alpha_{\mathrm{K}}}\frac{T_{\mathrm{K}}}{T_{\mathrm{Dy}}}\right)^{-3/2}\ .
\label{equa2}
\end{equation}
Based on the ratios of polarizabilities, masses and temperatures, we define the quantity
\begin{equation}
\label{A}
A= \left(1+\frac{m_{\mathrm{Dy}}}{m_{\mathrm{K}}}\frac{T_{\mathrm{K}}}{T_{\mathrm{Dy}}}\right)^{1/2}\left(1+\frac{\alpha_{\mathrm{Dy}}}{\alpha_{\mathrm{K}}}\frac{T_{\mathrm{K}}}{T_{\mathrm{Dy}}}\right)^{-3/2}\ .
\end{equation}
Combining (\ref{equa1},-\ref{A}) we deduce the collision rate
\begin{equation}
\Gamma_{\mathrm{coll}}=\sigma_{\mathrm{rel}}\frac{\displaystyle m_{\mathrm{Dy}}\bar{\omega}_{\mathrm{Dy}}^{3}}{\displaystyle \pi^{2}k_{\mathrm{B}}T_{\mathrm{Dy}}}N_{\mathrm{Dy}}N_{\mathrm{K}}A\ .
\end{equation}
We assume \cite{Mosk2001mou} that the average energy transfer during one Dy-K collision is given by $\Delta E=\xi k_{\mathrm{B}}(T_{\mathrm{K}}-T_{\mathrm{Dy}})$, where $\xi=4m_{\mathrm{Dy}}m_{\mathrm{K}}/(m_{\mathrm{Dy}}+m_{\mathrm{K}})^{2}$ accounts for the effect of the mass imbalance. Using this expression, we obtain the relation relating the collision rate $\Gamma_{\mathrm{coll}}$ and the thermalization time $\tau$
\begin{equation}
\tau^{-1}=\frac{1}{\Delta T}\frac{\mathrm{d}}{\mathrm{dt}}\Delta T=\xi\frac{N_{\mathrm{Dy}}+N_{\mathrm{K}}}{3N_{\mathrm{Dy}}N_{\mathrm{K}}}\Gamma_{\mathrm{coll}}
\end{equation}
One finally finds the relation
\begin{equation}
\tau^{-1}=\sigma_{\mathrm{el}}\;\frac{\xi}{3\pi^{2}}\frac{m_{\mathrm{Dy}}\bar{\omega}^{3}_{\mathrm{Dy}}}{k_{\mathrm{B}}T_{\mathrm{Dy}}}\left(N_{\mathrm{Dy}}+N_{\mathrm{K}}\right) A 
\end{equation}
between the thermalization time and the scattering cross section. Given the mass and polarizability ratios of Dy and K~\cite{Ravensbergen2018ado}, there is only a weak dependence of $A$ on the temperature ratio for the range of temperatures that we consider. We calculate $A\approx 1.5$, which allows us to analyze the near-exponential thermalization curve presented in Fig.~\ref{thermalization}.

\end{document}